\documentstyle{article}

\def\bb{\begin{equation}}
\def\ee{\end{equation}}
\def\th#1{\vspace{1mm}\noindent{\bf #1}\hspace{2mm}}

\begin{document}
\title{Symmetry approach in boundary value problems}
\author{I.T. Habibullin                            \\
        Ufa Institute of Mathematics, Russian Academy of Sciences,\\
        Chernyshevsky str. 112, 450000 Ufa, Russia                \\
        e-mail: ihabib@nkc.bashkiria.su}
\date{June 27, 1995}
\maketitle
\thispagestyle{empty}
\begin{abstract}
The problem of construction of the boundary conditions for nonlinear
equations is considered compatible with their higher symmetries.  Boundary
conditions for the sine-Gordon, Jiber-Shabat and KdV equations
are discussed. New examples are found for the JS equation.
\end{abstract}
\section{Introduction}
The subject of applications of classical Lie symmetries to boundary value
problems is well studied (see the monography \cite{ovs}). In
contrast, the question of involving higher symmetries to the same problem has
received much less attention, unlike say the Cauchy problem.  However, one
should stress that nowadays the higher symmetries' approach becomes the basis
of the modern integrability theory \cite{mss}.
A number of attempts to apply the inverse scattering method (ISM) to the
initial boundary value problem has been undertaken. It turned out that if
both initial data and boundary value are chosen arbitrary then the ISM loses
essentially its efficiency. On the other hand side the investigation by
E.Sklyanin \cite{skl} based on the $R-$ matrix approach demonstrated that
there is a kind of boundary conditions, compatible completely with the
integrability. The analytical aspects of such kind problems were studied in
\cite{tar}, \cite{bikt}. After \cite{hab1} it becomes clear that boundary
value problems found can effectively be investigated with the help of the
B\"acklund transformation.

Below we will discuss a higher symmetry test, proposed in \cite{hab2},
\cite{gur} to verify whether the boundary condition given is compatible with
the integrability property of the equation.  It is worthwhile to note that
all known classes of boundary conditions, compatible with integrability occur
to pass this symmetry test.  Boundary conditions involving explicit time
dependance for the Toda lattice compatible with higher symmetries has
recently been studied in \cite{adl}. It was established there that finite
dimensional systems obtained from the Toda lattice by imposing at both ends
boundary conditions consistent with symmetries were nothing else but
Painlev\'e type equations.

Let us consider the evolution type equation
\begin{equation}
u_{t}=f(u,u_{1},u_{2},...,u_{n})                        \label{eq}
\end{equation}
and a boundary condition of the form
\begin{equation}
p(u,u_{1},u_{2},...,u_{k}) \vert_{x=0}=0,               \label{gu}
\end{equation}
imposed at the point $x=0$. Here $u_i$ stands for the partial derivative of
the order  $i$ with respect to the variable $x.$ Suppose that the equation
given possesses a higher symmetry
\begin{equation}
u_{\tau}=g(u,u_{1},...,u_{m}).                          \label{sy}
\end{equation}
We call the problem (\ref{eq})-(\ref{gu}) compatible with the symmetry
(\ref{sy}) if for any initial data prescribed at the point $t=0$ a common
solution to the equations (\ref{eq}), (\ref{sy}) exists satisfying the
boundary condition (\ref{gu}). Let us explain more exactly what we mean.
Evidently one can differentiate the constraint (\ref{gu}) only with respect
to the variables $t$ and $\tau,$ (but not respect to $x$). For instance, it
follows from (\ref{gu}) that

\begin{equation} \sum_{i=0}^{n} {\partial p \over
\partial u_i}(u_i)_{\tau}=0,                            \label{dgu}
\end{equation}
where one should replace $\tau$-derivatives by means of the equation
(\ref{sy}).  The boundary value problem (\ref{eq})-(\ref{gu}) be compatible
with the symmetry (\ref{sy}) if the equation (\ref{dgu}) holds identically by
means of the condition (\ref{gu}) and its consequences obtained by
differentiation with respect to $t.$

To formulate an effective criterion of compatibility of the boundary value
problem with a symmetry it's necessary to introduce some new set of dynamical
variables consisting of the vector  $v=(u,u_{1},u_{2}, ... u_{n-1})$ and its
$t$-derivatives: $v_{t}$, $v_{tt}$, ... .  Passing to this set of variables
allows one really to exclude the dependance on the variable $x.$
In terms of these variables the symmetry (\ref{sy}) and the constraint
(\ref{gu}) take the form
\begin{equation}
v_{\tau}=G\,(v,v_{t} ,
v_{tt} ,... {\partial^{m_1}v\over \partial t^{m_1}}),           \label{sy1}
\end{equation}
\begin{equation}
P(v, {\partial v\over
\partial t},...,{\partial^{k_1}v\over \partial t^{k_1}})=0.
                                                                \label{gu1}
\end{equation}
The following criterion of compatibility was established in (\cite{gur}).

\th{Theorem.} The boundary value problem (\ref{eq})-(\ref{gu}) is
compatible with the symmetry (\ref{sy}) if and only if the
differential connection (\ref{gu1}) is consistent with the system
(\ref{sy1}).

We call the boundary condition (\ref{gu}) compatible with the integrability
property of the equation (\ref{eq}), if the problem (\ref{eq})-(\ref{gu}) is
compatible with infinite series  of linearly independent higher order
symmetries.

The problem of the classification of integrable boundary conditions is solved
completely for the Burgers equation (see \cite{gur})

\begin{equation}
u_{t}=u_{2}+2\,u\,u_{1},                                        \label{bur}
\end{equation}
\th{Theorem.} If the boundary condition $p(u,u_{1})\vert_{x=0}=0$ is
compatible at least with one higher symmetry of the Burgers equation
(\ref{bur}) then it is compatible with all even order  homogeneous symmetries
and is of the form $c_1(u_{1}+u^{2})+c_2\,u+c_3=0.$

In the Burgers case the boundary conditions of the general form (\ref{gu})
can also be described completely with the help of the "recursion operator for
the boundary conditions" $L={\partial \over\partial x}+u,$ which acts on
the set of integrable boundary conditions (see \cite{svi}). For instance, the
boundary condition
$L(c_1(u_{1}+u^{2})+c_2\,u+c_3)=c_1(u_{2}+3uu_{1}+u^{3})+c_2(u_1+u^2)
+c_3\,u=0$  is also integrable.

Let us describe boundary value problems of the form
\bb
a(u,u_x)\vert_{x=0}=0                                           \label{sggu}
\ee
\bb
u_{tt}-u_{xx}+\sin u=0,                                         \label{sg}
\ee
for the sine-Gordon equation compatible with the third order symmetry.

As it is shown in \cite{jsh} the complete algebra of higher symmetries for
the equation (\ref{sg}) i.e. $u_{\xi\eta}=\sin u,$ where  $2\xi=x+t,$
$2\eta=x-t$ splits into the direct sum of two algebras consisting of
symmetries of equations
$
u_{\tau}=u_{\xi\xi\xi}+u_{\xi}^3/2,\,
u_{\tau}=u_{\eta\eta\eta}+u_{\eta}^3/2,
$
correspondingly, which are nothing else but potentiated MKdV equation.
Particularly, the following flow commutes with the sine-Gordon equation
\bb
u_{\tau}=c_1(u_{\xi\xi\xi}+u_{\xi}^3/2)+c_2(u_{\eta\eta\eta}+
u_{\eta}^3/2).
                                                                \label{lcs}
\ee
The symmetry (\ref{lcs}) isn't compatible with any boundary condition of the
form (\ref{sggu}) unless $c_1=-c_2,$ under this constraint the  equation
(\ref{sggu}) is of one of the forms
\bb
u=const, \qquad
v=c_1\cos (u/2)+c_2\sin (u/2).                                  \label{sgco}
\ee
Note that the list of boundary conditions (\ref{sgco}) coincides with that
found by A.Zamolodchikov within the framework of the R matrix approach
\cite{zam}. The latter in (\ref{sgco}) in particular cases was studied
earlier in \cite{skl} and \cite{bikt}. The compatibility of the
former in (\ref{sgco}) with the usual version of ISM was declared earlier in
\cite{bikt}. But the statement was based in a mistake (see
\cite{hab3}). Our requirement of consistency is weaker than that is used
in \cite{bikt}.  Applications of these and similar problems for
the sine-Gordon equation and the affine Toda lattice in the quantum field
theory are studied in \cite{sal} and \cite{cor}.

According to the theorem above one reduces the problem of finding integrable
boundary conditions to the problem of looking for differential connections
admissible by the following system of equations, equivalent to (\ref{lcs})
with $c_1=-c_2$ and $v=u_x:$
\begin{equation}
\begin{array}{ll}
u_{\tau}=8u_{ttt}+6u_{t}\cos u+3v^2u_t+u_{t}^3,\\
v_{\tau}=8v_{ttt}+6v_{t}\cos u+6u_{tt}vu_t+3v^2v_t+3u_{t}^2v_t.\label{sgt2}
\end{array}
\end{equation}

One can prove that the boundary conditions (\ref{sgco}) are compatible with
rather large subclass of the sine-Gordon equation such that
\bb
u_{\tau}=\phi
(u,u_1,...u_{k_1})-\phi (u,\bar u_1,...\bar u_{k_1}),           \label{sgsym}
\ee
where $u_j=\partial^ju/\partial\xi^j,$ $\bar
u_j=\partial^ju/\partial\eta^j,$ and the equation
$u_{\tau}=\phi_i(u,u_1,...u_{k_i}),$ $i=1,2$ is a symmetry of the equation
$u_{\tau}=u_{\xi\xi\xi}+u_{\xi}^3/2.$

Another well-known integrable equation of hyperbolic type
\bb
u_{tt}-u_{xx}=\exp (u) +\exp (-2u)                              \label{ts}
\ee
has applications in geometry of surfaces. For the first time it was found by
Tzitzeica \cite{tzi}. The presence of higher symmetries for this equation
has been established by A.Jiber and A.Shabat \cite{jsh}. The simplest higher
symmetry of this equation is of the fifth order

\bb
u_{\tau}=u_{\xi\xi\xi\xi\xi}+5(u_{\xi\xi}u_{\xi\xi\xi}-u_{\xi}^2u_{\xi\xi\xi}
-u_{\xi}u_{\xi\xi}^2)+u^5_{\xi}.                                \label{tssym}
\ee
It is proved in the article cited that the symmetry algebra for (\ref{ts}) is
the direct sum of the symmetry algebras of (\ref{tssym}) and of the equation
obtained from (\ref{tssym}) by replacing $\xi$ by $\eta.$

Let us look for boundary conditions of the form
\bb
a(u,u_x)=0,                                                     \label{tsgu}
\ee
for the equation (\ref{ts}), compatible with the symmetry
\bb
\begin{array}{lr}
u_{\tau}=u_{\xi\xi\xi\xi\xi}+5(u_{\xi\xi}u_{\xi\xi\xi}-u_{\xi}^2u_{\xi\xi\xi}
-u_{\xi}u_{\xi\xi}^2)+u^5_{\xi}-u_{\eta\eta\eta\eta\eta}- \\
5(u_{\eta\eta}u_{\eta\eta\eta}-u_{\eta}^2u_{\eta\eta\eta}
-u_{\eta}u_{\eta\eta}^2)-u^5_{\eta}.                            \label{tst1}
\end{array}
\ee
Rather simple but tediously long computations lead to the following
statement.

\th{Theorem.} Boundary conditions (\ref{tsgu}) for the Jiber-Shabat equation
compatible with the symmetry (\ref{tst1}) (and then compatible with
integrability) are either of the form $u_x+c\exp (-u)\vert_{x=0}=0$ or
$u_x+c\exp (u/2)\pm \exp (-u) \vert_{x=0}=0 ,$ where $c$ is arbitrary.

Notice that all equations above are invariant under the reflection type
symmetry $x\rightarrow-x.$ It is unexpected that equations which don't admit
any reflection symmetry admit nevertheless boundary conditions compatible
with integrability. For instance, the famous KdV equation

\bb
u_t=u_{xxx}+6u_xu
\ee
is consistent with the boundary condition
\bb
u=0\vert_{x=0},\, u_{xx}\vert_{x=0}=0.                  \label{kdvcon}
\ee
It implies immediately that the boundary value problem
$$
u_t=u_{xxx}+6u_xu,\qquad
u=0\vert_{x=0}
$$
with the Dirichlet type condition at the axis $x=0$ admits
an infinite dimensional set of "explicit" finite-gap solutions.


This work was partially
supported by Russian Foundation of Fundamental Researches (grant
93-011-165) and International Scientific Foundation (grant
RK-2000).


\end{document}